\documentclass{aa}  

\usepackage{graphicx}
\usepackage{txfonts}
\usepackage{siunitx}
\usepackage{natbib}
\usepackage{amsmath}
\usepackage{mathtools}
\usepackage[breaklinks=true]{hyperref} 
\hypersetup{colorlinks=true,urlcolor=blue,citecolor=blue,pdfborder= 0 0 0}
\bibpunct{(}{)}{;}{a}{}{,}    

\usepackage{caption}
\usepackage{subcaption}
\usepackage{color}
\definecolor{darkgreen}{rgb}{0.0,0.5,0.0}
\definecolor{darkred}{rgb}{0.7,0.,0.3}
\definecolor{purple}{rgb}{0.5,0,0.5}
\definecolor{gray}{rgb}{0.5,0.6,0.7}

\begin{document}

   \title{AXES-SDSS: Solving the puzzle of X-ray emission of optical galaxy groups via a modified Hausdorff distance} 
\titlerunning{MHD for AXES-SDSS} 
   \author{J. Kosowski
          \inst{1}
          \and
          A. Finoguenov\inst{1}
          \and
          E. Tempel\inst{2,3}
          \and
          G. A. Mamon\inst{4}
          }

   \institute{Department of Physics, University of Helsinki, , Gustaf Hällströmin katu 2A, Helsinki, FI-00014, Finland 
         \and
   Tartu Observatory, University of Tartu, Observatooriumi 1, 61602 T\~oravere, Estonia
\and
Estonian Academy of Sciences, Kohtu 6, 10130 Tallinn, Estonia
\and
Institut d’Astrophysique de Paris (UMR 7095: CNRS \& Sorbonne Université), F-75014 Paris, France             
             }

   \date{Received December 16 2024; accepted }

 
  \abstract
   {
    The identification of X-ray and CMB sources as galaxy groups and clusters is a prerequisite for cluster cosmology. But the identification of groups, especially nearby ones, suffers from projection effects which in turn affect the purity of the sample.
   }
   {
   In X-rays, the position of the cluster can be given either by the peak of the emission, or by the full information content of the cluster image. Similarly, the optical center, or its member galaxies, can describe the optical counterpart. With the progress of numerical simulations, it is currently feasible to reproduce both the optical group membership assignment and the behavior of the group outskirts in X-rays, and therefore there is an opportunity to define a reproducible group identification procedure.
   }
   {
   We performed two-way matching between X-ray contours, drawn at a fixed surface brightness level corresponding to a baryonic overdensity of $\sim$500, to the projected galaxy positions using a modified Hausdorff distance (MHD). We used the volume-limited SDSS group catalog to evaluate the purity and completeness of the procedure, maintaining the constant performance of the optical group finder with redshift.
   }
    {
    We find that an MHD of 0.631 Mpc provides 90\% purity. This is a clear improvement over the methods that rely on the optical counterparts' distance and richness.
    We study the purity versus MHD and the completeness versus redshift and velocity dispersion. Over half of nearby groups have X-ray emission, even those with velocity dispersions as low as 200$\,\mathrm{km}\,\mathrm{s}^{-1}$; this has never previously been demonstrated. The bulk of these groups follow the same scaling relations as the groups with a small separation between the optical and X-ray centers, removing feedback as an explanation for the lack of matches in previous studies. Instead, the problem is caused by over-merging in the optical group catalog construction and source confusion in X-rays.
    }
   {}


   \maketitle
%

\section{Introduction}

Studies of galaxy clusters play an important role in cosmology, delivering competitive constraints associated with the growth of structure in the Universe in late epochs \citep[for a recent review see][]{ClercFinoguenov}. Achieving ever-increasing goals for the precision of cosmological studies requires improvements to be made to all aspects of cluster studies. The identification of galaxy clusters is a prerequisite for any study and has to be validated. One successful approach is to use an external cluster scoring parameter, such as richness, to control the purity of cluster identification and to model completeness \citep{Finoguenov20}.  
Projection effects strongly affect the identification performance at $z<0.2$, and hydrodynamical effects lead to complications with using the centers of emission for identification. In addition, selecting cluster centers based on the most massive galaxy introduces a bias toward centering on infalling groups, as there is little relation between of mass of the central galaxy and the mass of the system at high halo masses. 

The best way of validating the method is to use the results of numerical simulations, which provide the mock sky data. The full modeling of galaxy clusters still suffers today from uncertainties about the role of cooling and feedback, which affect the physics of cluster cores \citep{300cl}. Good progress has been made, though, in modeling the cluster outskirts, establishing a link between the baryonic fraction and the modifications of the shear power spectrum \citep{vanDaalen20}. Thus, a consistent set of map-making and cosmological analysis is possible. At the same time, galaxy mocks have reached the stage of reproducing local galaxy surveys well, and spectroscopic group selection is well under control \citep{2014MNRAS.441.1513O, 2018MNRAS.475..853O, uchuusdss}. 
These developments highlight the importance of group identification methods that rely on the part easiest to model: the group outskirts.

In this paper, we support this approach by considering the identification of cluster outskirts. We develop a practical application of the method to the identification of large X-ray sources, found in ROSAT All-Sky survey (RASS), using the optical group catalog of SDSS. These datasets are currently best suited for this work, with RASS uniquely providing the information on X-ray emission on scales of about one degree, and SDSS being the largest survey of nearby optical groups that matches the sensitivity of RASS data.

In performing the identification, we consider the advantages offered by the peculiar spatial size of nearby objects, which makes them very distinct from the distant clusters. To accomplish this, we employ a modified Hausdorff distance (MHD), which is a mathematical construct that we see fit for our purposes.

The paper is structured as follows. In Sect. 2, we describe the SDSS optical group catalog. In Sect. 3, we describe the catalog of RASS X-ray sources. In Sect. 4, we outline the identification method we propose to use. In Sect. 5, we present the results of our analysis on the completeness and purity of the procedure. We conclude in Sect. 6. 

For conversions between angular and physical separations, we use the WMAP7 cosmological parameters as defined in \cite{komatsu2011WMAP7}. These parameters are used to maintain consistency with the optical group catalog described in Sect. 2.

\section{SDSS optical group catalog}

We used the optical friends-of-friends (FoF) group catalog by \citet{Tempel2014}, which is based on the SDSS Data Release 12 (DR12; \citet{Blanton2005,Adelman-McCarthy2008,Padmanabhan2008,Eisenstein2011, Alam2015}). The catalog covers the largest contiguous area of the main galaxy sample of the SDSS Legacy Survey. In contrast to \citet{Tempel2017}, which was used in our first study of AXES-SDSS \citep{Damsted24}, we used a set of volume-limited catalogs, each of which has a fixed linking length, set to approximately trace the virial extent of galaxy groups. These catalogs have uniform group definitions as a function of redshift, which simplifies the interpretation of our results and the choice of matching distance.

The galaxy redshifts were corrected for motion with respect to the cosmic microwave background and limited to $z<0.2$. We concentrated on using two catalogs: v180, which is complete down to $M_R=-18.0$ and provides the deepest spectroscopic group catalog for this study, and v195, which is complete to $M_R=-19.5$ and covers a larger redshift range, allowing us to study the redshift evolution of our matches and deduce the RASS selection function. Table 1 from \cite{Tempel2014} defines the exact redshift cuts (and provides additional properties) for each volume-limited catalog. Additionally, their Figure 2 showcases the relationships between magnitude, redshift, and comoving distance for the catalogs. For convenience, we have provided these key properties for the two catalogs relevant to our paper in Table \ref{tab:cat_summary}.

\begin{table}

    \caption{Volume-limited catalog properties}
    \label{tab:lx_sig_params}
    \centering
    \begin{tabular*}{0.45\textwidth}
    {l@{\extracolsep{\fill}}lccr}
        \hline\hline
        \multicolumn{1}{l}{Catalog} & \multicolumn{1}{c}{$M_{R,\rm lim}$} & \multicolumn{1}{c}{$z_{\rm lim}$} & \multicolumn{1}{c}{$d_{\rm lim}$}\\
         & mag &  & $h^{-1}$\,Mpc\\
        
        \hline
        \texttt{v180} & --18.0 & 0.045 & 135.0 \\
        \texttt{v195} & --19.5 & 0.089 & 261.3 \\

        \hline
    \end{tabular*}
    \label{tab:cat_summary}
\end{table}

As we need to reliably trace group outskirts, we set stringent requirements on the number of members being ten or more. However, as advances in optical surveys far supersede the advances in X-ray observations, we believe that such requirements will soon be met for the entire sky with advances of DESI \citep{desi} and 4MOST surveys \citep{2019Msngr.175....3D, 2023Msngr.190...46T}. To control the extent of group membership, we need catalogs produced using a fixed linking length, which in turn requires the volume-limited catalog of galaxies, complete down to some absolute magnitude. 

To deliver consistent results, we need to ensure that the matching is done to the same part of the group or cluster. Within the spectroscopic range covered by the SDSS groups, a strong deviation in the apparent size of systems occurs below a redshift of 0.01. Above this redshift, the angular scale adopted for the detection of X-ray emission covers the virial radius, while below it covers mainly the core, creating a mismatch with the group members, which still trace the virial scales of the group. This can also be understood as a property of wavelet decomposition, which reports the excess of the emission on the selected spatial scales, not the actual emission level. For nearby groups, this is higher than what is traced by our contours. On the other hand, the main advantage of the SDSS group catalog is to move beyond a redshift of 0.025 and deliver a large set of groups. Thus, we selected $z>0.025$ as our minimum redshift. The upper limit on the redshift of this study was set by the sensitivity of both RASS and SDSS surveys, which leaves the group regime at a redshift of 0.1.

\section{AXES: Large X-ray sources in RASS}

In this work, we have used a new reprocessing of ROSAT All Sky Survey (RASS) data.  We combined the wavelet scales of 12 and 24 arcmin after removing the emission detected on scales of 6 arcmin and below. Further details on the wavelet decomposition can be found in \cite[][\footnote{\url {https://github.com/avikhlinin/wvdecomp}}]{vikhlinin98}. 
\cite{Khalil24} study the properties of the X-ray sources detected using this method, noticing a large spread of radii of source detection. To homogenize the source extent, in this paper, we took a different approach to X-ray source characterization. We extracted X-ray contours using a sum of wavelet maps on scales of 12 and 24 arcminutes. This ignored the parts of the detection that extend to larger scales, important for $z<0.01$, which are therefore not covered by this study. Given the limited redshift range of our study, the effect of surface brightness dimming is moderate and the selected contour level corresponds to the level of equal baryonic overdensity. At the same time, the construction of the optical membership was also done to ensure an approximately similar tracing of overdensity. Thus, there is an opportunity to tune the description of the X-ray emission to match the optical group definition. 
By experimenting with the contour levels, we selected the contour level value of $2 \times 10^{-5}$ counts per second and pixel in the wavelet maps on spatial scales of 12--24 arcminutes created using RASS images in the 0.5--2 keV band, which for extragalactic field \ion{H}{i} column density ($\mathrm{N_H}$) conditions translates into surface brightness of $\rm 1.6 \times 10^{-12}\, erg\,s^{-1}\,cm^{-2}\,deg^{-2}$. The contours drawn at this level match well the extent of X-ray sources in \cite{Khalil24}. 
Compared to the selected threshold for X-ray contours, lower values of surface brightness lead to increased source confusion on the spatial scales of our analysis. Higher levels would miss galaxy groups. 
Ultimately, the effect of the choice of surface brightness level will be modeled with current state-of-the-art hydrodynamical simulations. 
The chosen level of X-ray surface brightness levels encloses an area of 1266 square degrees in the all-sky RASS maps. This is at a boundary of the source confusion, defined as a ratio of the sky area to the source area to be larger than 40. This implies that deeper X-ray data would be confused on the spatial scales we study and that we would have to study smaller angular scales of source emission. The situation is only 20\% better if we concentrate on the extragalactic sky (defined at X-rays as $\mathrm{N_H}<7\times 10^{20}$ cm$^{-2}$, a condition achieved over 60\% of the sky), where the X-ray emission enclosed within the contours of the selected level occupy 624 square degrees.

\section{Outline of the identification method}

\subsection{Modified Hausdorff distance}

We based our identifications on a modified form of the Hausdorff distance (HD) method of calculating distances between datasets. The standard HD ($d_{\mathrm{H}}(X,Y)$) for two finite point sets, $X=\{x_1,x_2,...x_i\}$ and $Y=\{y_1,y_2,...y_i\}$, is shown in equation \ref{eq:HD}, {with $d(X,Y)$ being the “directed” HD as defined by equation \ref{eq:dHD}} (\cite{Huttenlocher1993}).
\begin{equation}
        d_{\mathrm{H}}(X,Y):=\max\left[d(X,Y),d(Y,X)\right]
    \label{eq:HD}
\end{equation}

\begin{equation}
    d(X,Y)=\max_{x\in X}\left[\min_{y\in Y}\lVert x-y\rVert\right]
    \label{eq:dHD}
\end{equation}
In other words, $d_\mathrm{H}$ is the largest distance of a point in a set to the nearest point in the other set. With this definition, the HD can be used as an indicator of how close two sets are from each other in a given metric space. \citeauthor{Huttenlocher1993} first used this metric for arbitrary shape matching in computer vision in 1993.

However, the use of the maximum within the directed HD makes the final HD susceptible to drift from outliers. To solve this, \cite{Dubuisson1994} proposed a MHD that uses the mean instead of the maximum for the directed HD and was able to prove its efficacy. \cite{Montes2019} applied this MHD to an astrophysical context by using the MHD to quantify similarities between distributions of mass, intracluster light (ICL), and hot X-ray emissions within galaxy clusters. However, this study uses contour maps for both sets compared using MHD, where we use a set of contours and a set a points. With this in mind, we decided to further modify the MHD.

The mathematical beauty of using a maximum difference is spoiled by the astrophysical reality in which contamination plays an important role. The optical membership is affected by interlopers, while X-ray contours can be affected by another X-ray source. So, we have replaced the categorical maximum within the {directed HD with the median, which implies that we consider most of our contour points and most of the galaxies to be associated with a cluster, but not all. By using the median specifically, we calculated the distances based on the best 50\% of group members and contour points. In addition, we replaced the maximum of the two directed HDs with their mean to mitigate any bias the maximum may have toward either set due to their different structures (contours vs. points). With this, we have our new MHD (Eqs.~\ref{eq:mhd} and \ref{eq:mhd_dir}).}

\begin{equation}
    d_{\mathrm{H,mod}}(X,Y):=\mathrm{mean}\left[d_{\mathrm{mod}}(X,Y),d_{\mathrm{mod}}(Y,X)\right] \ ,
    \label{eq:mhd}
\end{equation}

\begin{equation}
    d_{\mathrm{mod}}(X,Y)=\underset{x\in X}{\mathrm{med}} \left[\min_{y\in Y}\lVert x-y\rVert\right] \ .
    \label{eq:mhd_dir}
\end{equation}

\subsection{Establishing optical/X-ray matches}

For the final implementation of this method, we did not simply run the MHD algorithm across the sets of optical and X-ray data. Instead, one set of median closest distances was calculated using the optical data and the contours extracted from the X-ray sources, with respect to the contours; the other set was calculated using the optical data and the nonzero pixels of the wavelet reconstruction of the X-ray image, combining the wavelet scales in 12, and 24 arcminutes. The average of these two sets was our final MHD value for that optical/X-ray source combination. This approach allowed us to ensure that our ideal optical groups overlapped the X-ray source in question, but still contributed to the structure of the contour.

\begin{figure}
\centering
\begin{subfigure}{.35\textwidth}
    \centering
    \includegraphics[width=.95\linewidth]{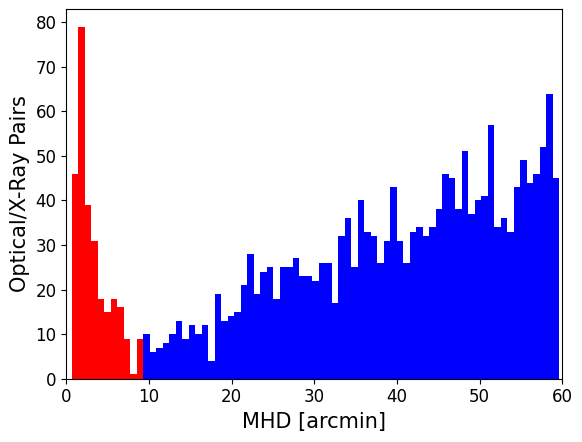}  
    \caption{}
    \label{fig:mhd_hist_deg_v195}
\end{subfigure}
\begin{subfigure}{.35\textwidth}
    \centering
    \includegraphics[width=.95\linewidth]{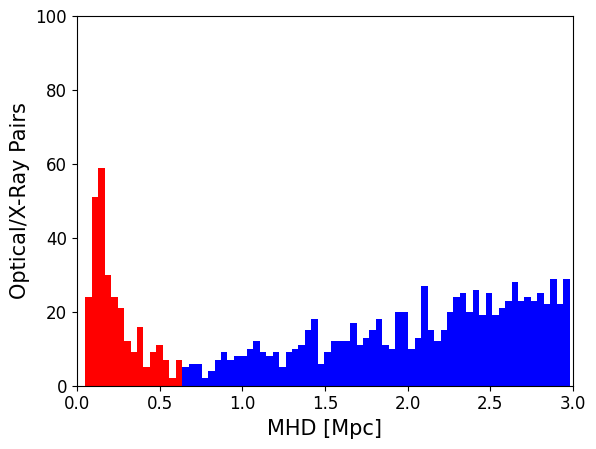}  
    \caption{}
    \label{fig:mhd_hist_mpc_v195}
\end{subfigure}
\caption{Number of optical to X-ray combinations as a function of their corresponding MHD value. {\it (a)} the MHD was calculated in angular units; {\it (b)} the MHD was calculated in physical units, using the redshift of the matched optical group. For each figure, the red region highlights the combinations that fall within the $5\%$ confusion limit and are therefore considered matched optical/X-ray pairs.}
\label{fig:mhd_hist_v195}
\end{figure}

Using equation \ref{eq:mhd}, we calculated MHDs for all combinations of optical galaxy groups and X-ray contours extracted from the RASS sources. The procedure results in two MHD values, one in angular units and one in physical units. The MHDs are used separately and lead to two different sets of results: matches defined by the MHD limit in degrees and matches defined by MHD limit in megaparsecs. Figure~\ref{fig:mhd_hist_v195} shows the distribution of each of these distances in units of arcminutes and megaparsecs. A roughly linear trend is seen for an MHD of $>9.54\,\,\mathrm{arcmin}$ and $>0.631\,\,\mathrm{Mpc}$ (in blue) caused by random matches at large distances. 
Below this threshold, an overdensity {(in red)} in the number of matches is distinguishable. The greater edge of this overdensity represents the smallest distance at which real optical/X-ray matches become more significant than random matches. For this upper MHD limit of real matches, we used the weighted 5\% confusion limit of matches, scaled to the ratio of total optical groups to X-ray contours of our sample. {This limit was determined by first defining the scale ratio of the number of unique optical groups to unique X-ray contours. Multiplying this ratio by 5\% gave us our weighted confusion limit. We then iterated over our potential matches to determine which maximum MHD value produces the same confusion limit. That value was used as the maximum MHD limit.}

\begin{figure}
\centering
\begin{subfigure}{.25\textwidth}
    \centering
    \includegraphics[width=.95\linewidth]{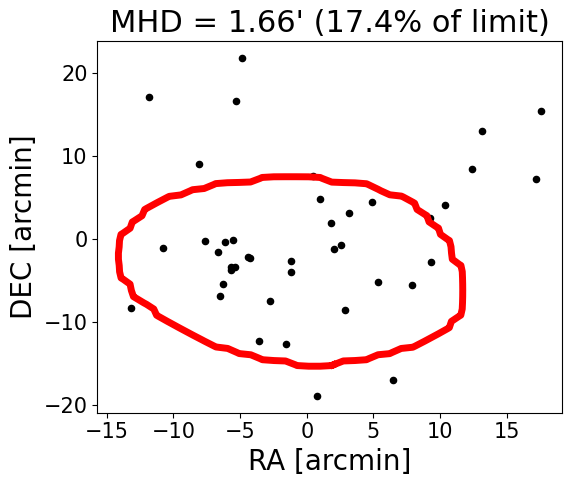}  
    \caption{}
    \label{fig:match}
\end{subfigure}%
\begin{subfigure}{.25\textwidth}
    \centering
    \includegraphics[width=.95\linewidth]{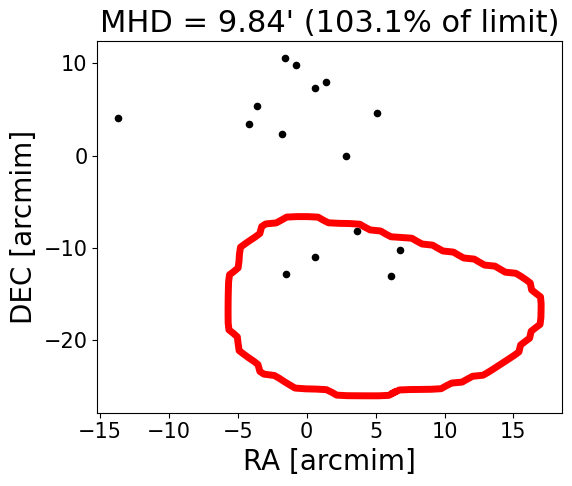}  
    \caption{}
    \label{fig:nonmatch}
\end{subfigure}
\caption{Examples of both a matched (\ref{fig:match}) and an unrelated (\ref{fig:nonmatch}) pair of X-ray (red contour) and optical (black dots) group components, determined by the MHD calculations. The maximum MHD allowed for a match is {9.54 arcmin}.}
\label{fig:match_example}
\end{figure}

Figure~\ref{fig:match_example} shows examples of a match and a non-match, according to our method. The left panel shows an optical/X-ray pairing with an MHD of 1.66 arcmin, well below the upper limit of 9.84 arcmin. By eye, this pairing looks fairly good, with a large overlap between contour and optical members and with the contour aligned with the region of highest member density. Conversely, the right panel shows an optical/X-ray pairing that our method determines to be due to random chance, with an MHD of 0.658 Mpc. While there is some overlap in the X-ray emission and group members, which could imply a connection, the large-scale association between the two components is not high enough for a confident match.

\begin{figure}
  \centering
    \includegraphics[width=0.5\textwidth]{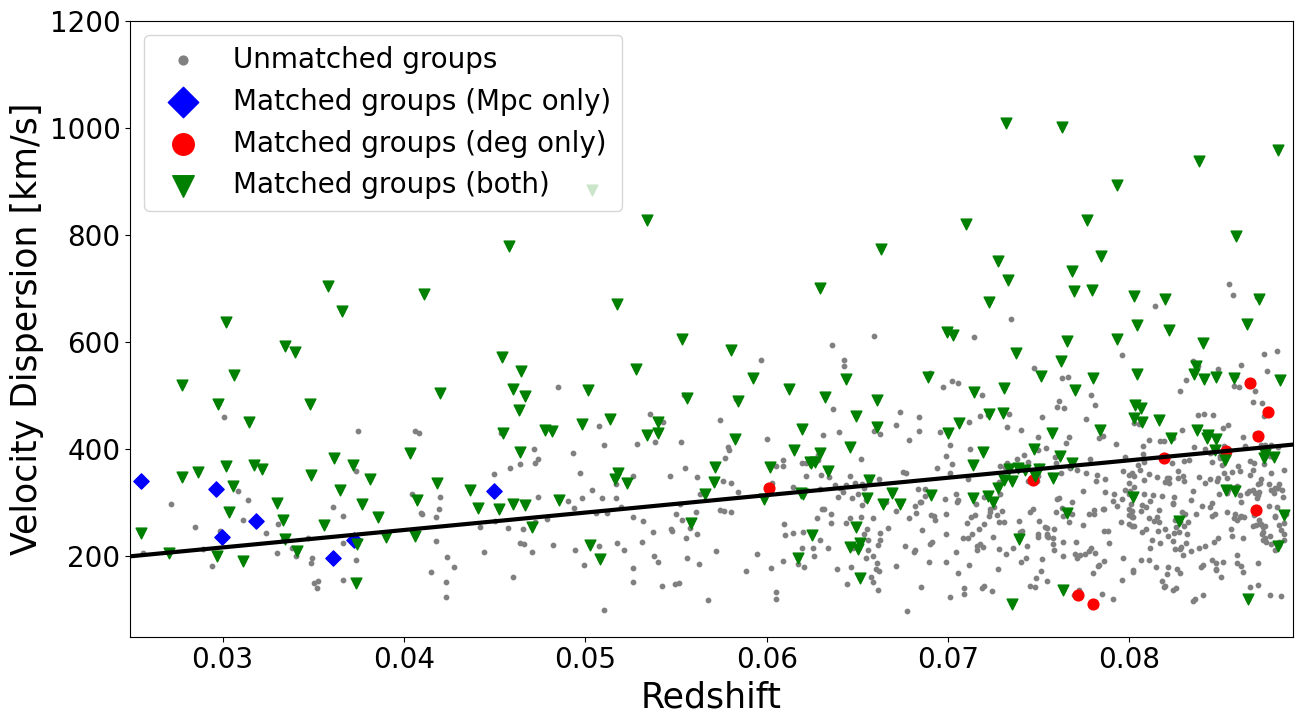}
  \caption{Identification of X-ray emission with galaxy groups from the v195 volume-limited catalog of \cite{Tempel2014}. The sample was cleaned to remove groups with fewer than ten member galaxies. Groups that were only matched with an X-ray source using an MHD in degrees are marked with red circles, while groups that were only matched using an MHD in megaparsecs are marked with blue diamonds. Groups that were matched when using an both an MHD in megaparsecs or degrees are marked with green triangles. Unmatched groups are plotted in gray. The black line represents a $\sigma (z)$ function above which we have 50\% completeness for matches.}
  \label{fig:velD_vs_Z_195}
\end{figure}

Using the MHD limit, we can begin to estimate the exact number of matches of groups. Figure~\ref{fig:velD_vs_Z_195} shows that we gain potential matches across the full sample. Notably, we see a majority of groups with matches in the low-z, low-$\sigma$ regime. We see a drop in the sensitivity of the group detection as a function of the redshift from 200 $\mathrm{km}\,\mathrm{s}^{-1}$ at $z=0.025$ to 400$\,\mathrm{km}\,\mathrm{s}^{-1}$ at $z=0.09$ at the $50\%$ completeness level. This evolution is faster than expected if the detection was limited by surface brightness dimming (e.g., we would expect $\sim 5\%$ increase over same velocity dispersion range), and therefore the detections are flux-limited. The increased drop in sensitivity is likely driven by the combined effect of the small angular size of the X-ray emission of groups at high redshift, detection based on the area integral of surface brightness, and photon-limited detection on small spatial scales \citep[for a detailed study, see][]{seppi25}.

\section{Results}

\begin{figure*}
\centering
\begin{subfigure}{.45\textwidth}
    \centering
    \includegraphics[width=0.9\linewidth]{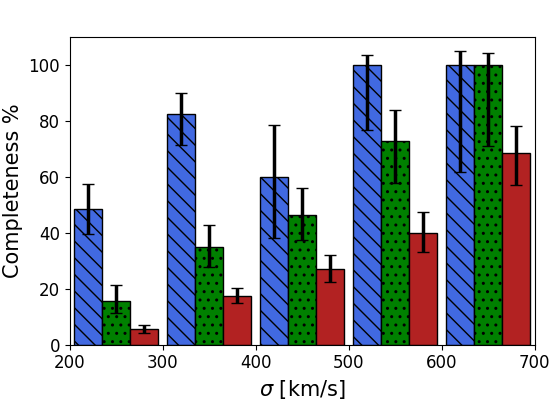}  
    \caption{}
    \label{fig:completeness_purity_Mpc_comp}
\end{subfigure}
\begin{subfigure}{.45\textwidth}
    \centering
    \includegraphics[width=0.9\linewidth]{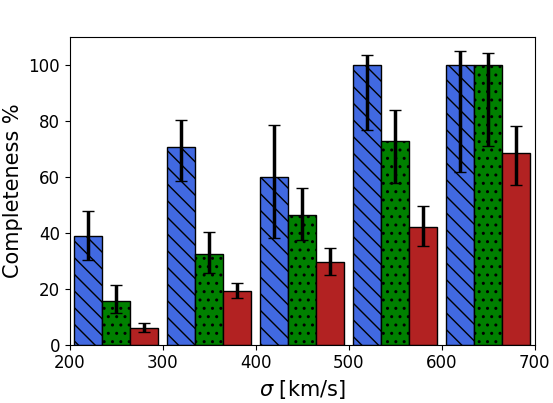}  
    \caption{}
    \label{fig:completeness_purity_Mpc_pur}
\end{subfigure}
\begin{subfigure}{.45\textwidth}
    \centering
    \includegraphics[width=0.9\linewidth]{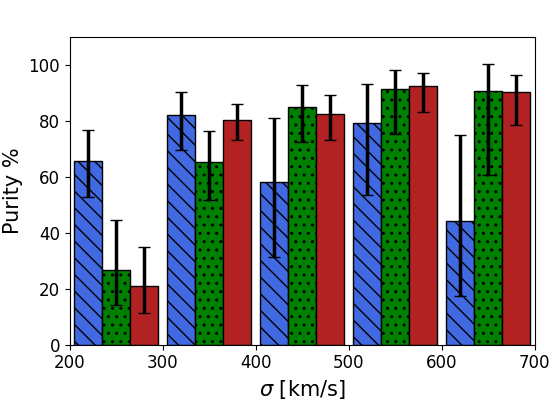}  
    \caption{}
    \label{fig:completeness_purity_deg_comp}
\end{subfigure}
\begin{subfigure}{.45\textwidth}
    \centering
    \includegraphics[width=0.9\linewidth]{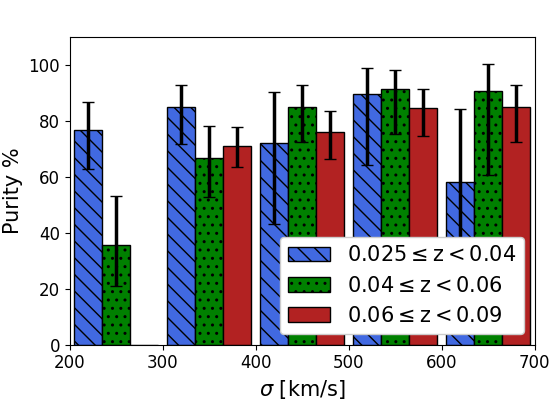}  
    \caption{}
    \label{fig:completeness_purity_deg_pur}
\end{subfigure}
  \caption{Completeness (top panels) and purity (bottom panels) of the optical/X-ray matches identified by applying our MHD identification method to the optical galaxies v195 volume-limited catalog of \cite{Tempel2014} and X-ray contours from RASS. Completeness and purity are presented as a function of bins of velocity dispersion of 100$\,\mathrm{km}\,\mathrm{s}^{-1}$. The data was also split into three bins of redshift across the z range of the v195 catalog and the legend on panel (d) applies to all panels. Completeness and purity are shown for each bin of redshift and velocity dispersion. Left column: Completeness and purity for matches determined by an MHD calculated in megaparsecs, based on the redshift of the matching optical galaxy group. Our MHD method excels at identifying low-z groups even for $\sigma$ below 300$\,\mathrm{km}\,\mathrm{s}^{-1}$. Right column:  Completeness and purity for matches determined by an MHD calculated in degrees. The completeness of matches is generally equal or lower when using degrees rather than physical units.}
  \label{fig:completeness_purity}
\end{figure*}

To quantify the quality of the optical/X-ray source matches, purity and completeness were calculated. In computing the completeness of the data, we need to establish the number of true matches contained in the data. This is done by comparing the excess match probability with the chance one. These numbers are lower than the full optical catalog for a number of reasons, primarily dependent on the X-ray completeness as well as the purity of the optical catalogs. Using these numbers we can discuss the correspondence between the X-ray and optical groups. {The final completeness was determined as the ratio of chance matches to the total potential matches (chance and excess matches).}
The purity of the catalog was determined by first creating a scrambled set of galaxy groups. This set would be constructed from the real groups the v195 catalog, but with random offsets applied to each group. This way, we can maintain the relative galaxy positions within each group while having a random group distribution. We calculated random matches from this set and the purity was calculated as one minus the ratio of random matches to true matches (ones found using the unscrambled set) for each bin. A large number of low-mass groups quickly reduces the purity of the match with increasing distance, even when considering the physical scale of the group.

\begin{figure*}
\centering
\begin{subfigure}{.45\textwidth}
    \centering
    \includegraphics[width=0.9\linewidth]{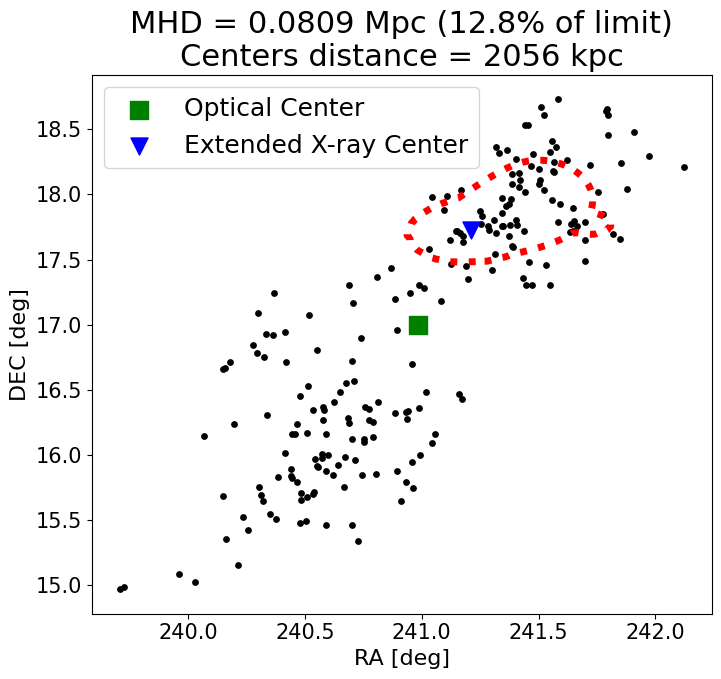}  
    \caption{}
    \label{fig:centers_examples1}
\end{subfigure}
\begin{subfigure}{.45\textwidth}
    \centering
    \includegraphics[width=0.9\linewidth]{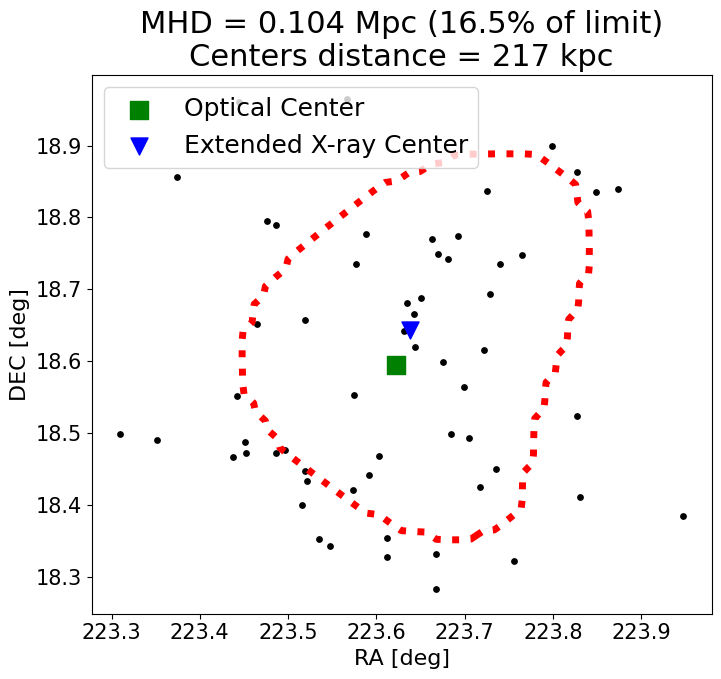}  
    \caption{}
    \label{fig:centers_examples2}
\end{subfigure}
\begin{subfigure}{.45\textwidth}
    \centering
    \includegraphics[width=0.9\linewidth]{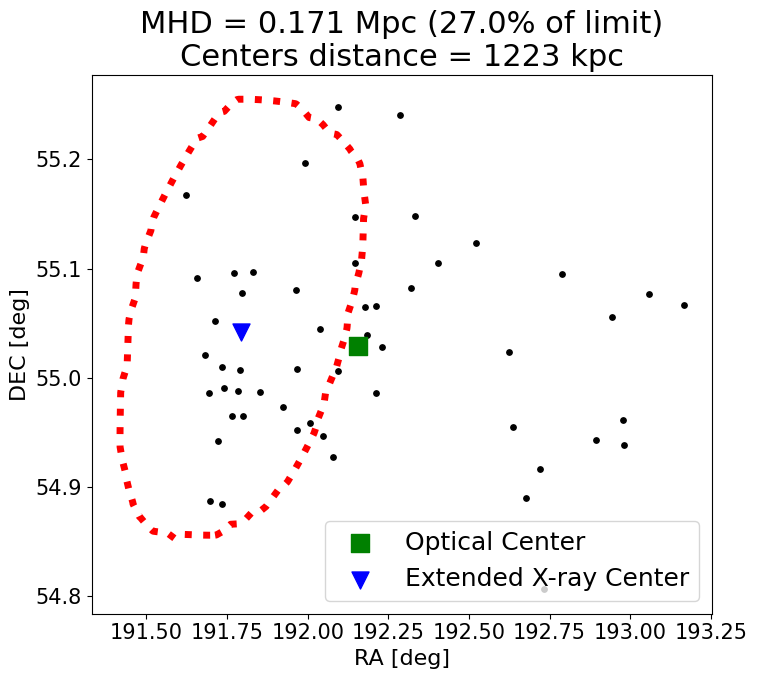}  
    \caption{}
    \label{fig:centers_examples3}
\end{subfigure}
\begin{subfigure}{.45\textwidth}
    \centering
    \includegraphics[width=0.9\linewidth]{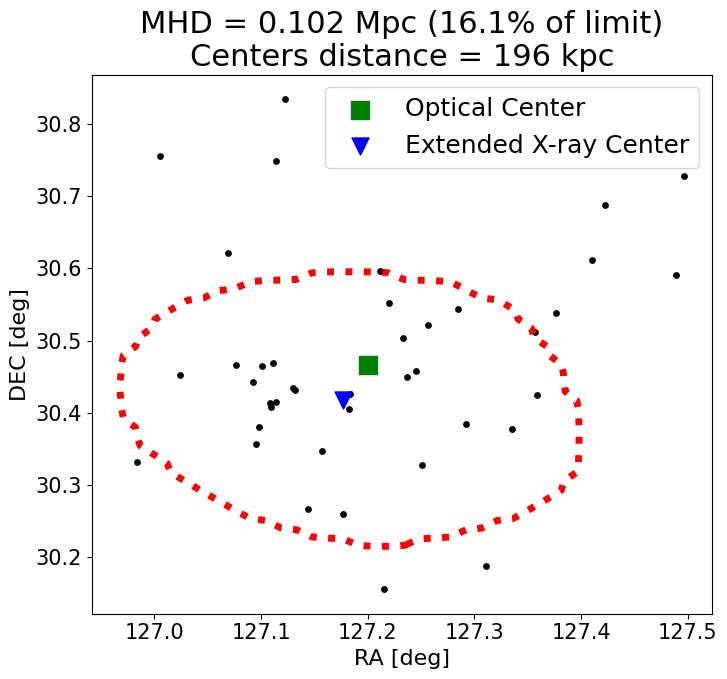}  
    \caption{}
    \label{fig:centers_examples4}
\end{subfigure}
  \caption{Four examples of group identifications with the optical and extended X-ray centers shown. Despite all of these examples exhibiting very confident match criteria, no strong correlation exists between the MHD value (i.e., chance of a match) and the distance between the optical and X-ray centers. In some cases, such as in figures \ref{fig:centers_examples1} and \ref{fig:centers_examples2}, the optical centers even lie outside of the X-ray contour, over 1000 kpc from the X-ray centers.}
  \label{fig:centers_examples}

\end{figure*}

Plots of the purity and completeness as a function of velocity dispersion ($\sigma$) and redshift are shown in figure~\ref{fig:completeness_purity}. The top panels show completeness as a function of $\sigma$ and are split into three redshift bins. The bottom panels show the purity, also as a function of $\sigma$, with the same redshift bins. Using the physical or angular MHD does affect the overall quality of the final catalog. Completeness is generally better when using the physical MHD and the purity is comparable to that of the angular MHD matches. We chose to focus on physical MHD for our final catalog as it provided a significant increase in completeness with minimal loss in purity. 

In calculating the fractions, $p$, (of $x$ successes having a total of $n$ trials) we used the Agresti-Coull method (Eq.~\ref{eq:bimodial error}, \cite{ac98}) to compute the 68\% confidence level interval ($z_{\alpha/2}=1$).

\begin{equation}
\widetilde{p}\pm z_{\alpha/2}\sqrt{\dfrac{\widetilde{p}(1-\widetilde{p})}{\widetilde{n}}}, \text{where } \widetilde{p}=\dfrac{x+{z_{\alpha/2}^2}/{2}}{n+z_{\alpha/2}^2} \text{and } \widetilde{n}=n+z_{\alpha/2}^2
\label{eq:bimodial error}
\end{equation}

We find that the MHD method works remarkably well for galaxy groups of low redshift and low $\sigma$. While completeness drops for bins of higher redshift, the purity remains above {60}\% for most bins of velocity dispersion. The 100\% completeness shown in the highest-velocity dispersion bin should not be taken at face value. Our MHD method was able to identify a match for all groups in that bin, but the very small numbers of groups present within that regime make it difficult to discern proper statistics. 

The relative lack of identifications in low-$\sigma$ bins (particularly in the $200-300\,\mathrm{km}\,\mathrm{s}^{-1}$ bin) can potentially help us understand the lack of detection of X-ray emission for low-mass galaxy groups. The limits of the data are such that the exploratory territory is the rich groups with velocity dispersions in the $200-500\,\mathrm{km}\,\mathrm{s}^{-1}$ range. Lower-mass systems would be largely incomplete due to RASS sensitivity, while cluster identification is not so challenging that it would require substantial improvement. Thus, it is of special interest to define the route to a successful selection for this category.

Analyzing the distance between the X-ray and optical centers of matched pairs shows clear deviations from the assumption that large separations between centers signify no associations. Figure \ref{fig:centers_examples} showcases four matched groups and their respective X-ray and optical centers. All four matches are well below the upper MHD limit but have very different separations between centers. Figures \ref{fig:centers_examples1} and \ref{fig:centers_examples3} are most notable as the optical centers lie outside the X-ray contour and have separations of 2056 kpc and 1223 kpc, respectively. These groups may have been deemed to be unassociated based on the centers alone. Regardless, the MHD method is able to use the large-scale structure of both the X-ray and optical components to identify a probable match. {Furthermore, the group in figure \ref{fig:centers_examples1} appears to show two separate galaxy concentrations, which could imply that this group is a result of over-merging and is in fact two separate groups. It is notable that the MHD method was still able to identify an association with a potentially over-merged group.}

\begin{figure*}
  \centering
\begin{subfigure}{.425\textwidth}
    \centering
    \includegraphics[width=.95\linewidth]{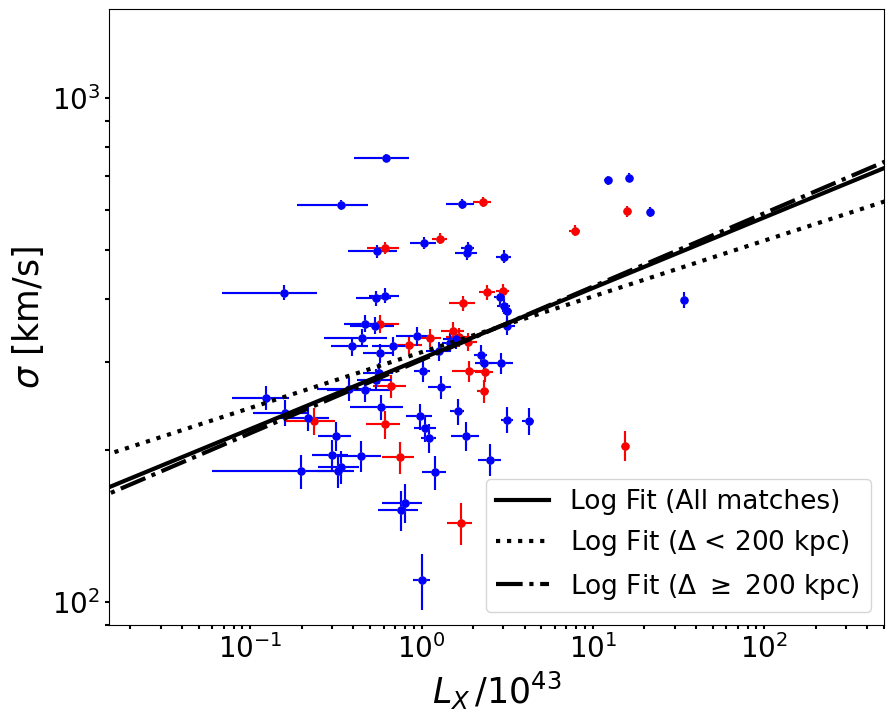}  
    \caption{}
    \label{fig:velD_Lx_180}
\end{subfigure}
\begin{subfigure}{.35\textwidth}
    \centering
    \includegraphics[width=.95\linewidth]{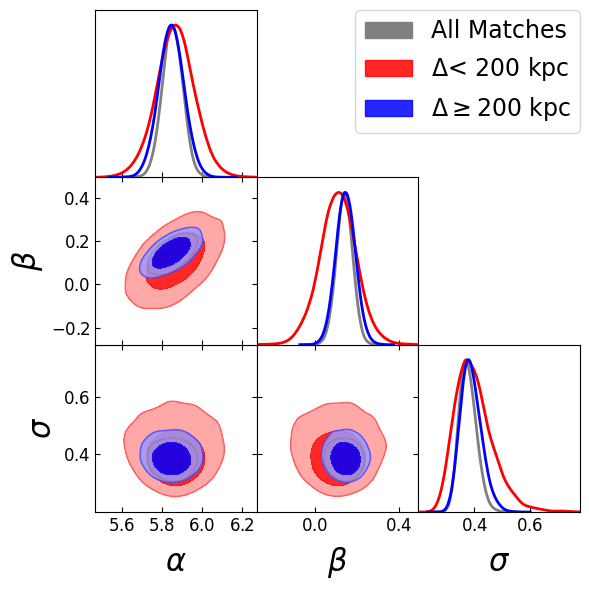}  
    \caption{}
    \label{fig:triangle_plot_180}
\end{subfigure}
\begin{subfigure}{.425\textwidth}
    \centering
    \includegraphics[width=.95\linewidth]{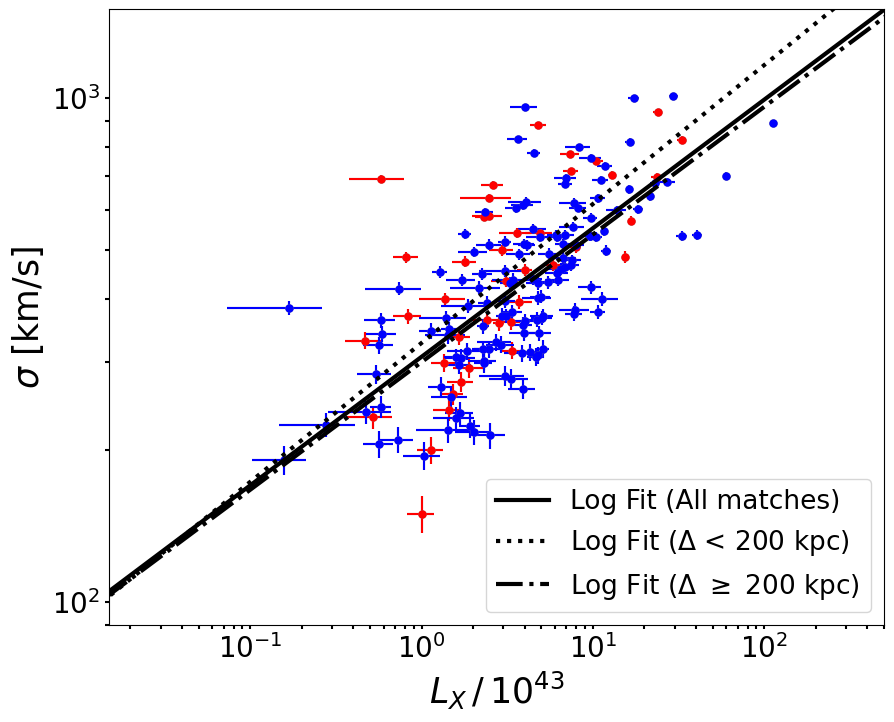}  
    \caption{}
    \label{fig:velD_Lx_195}
\end{subfigure}
\begin{subfigure}{.35\textwidth}
    \centering
    \includegraphics[width=.95\linewidth]{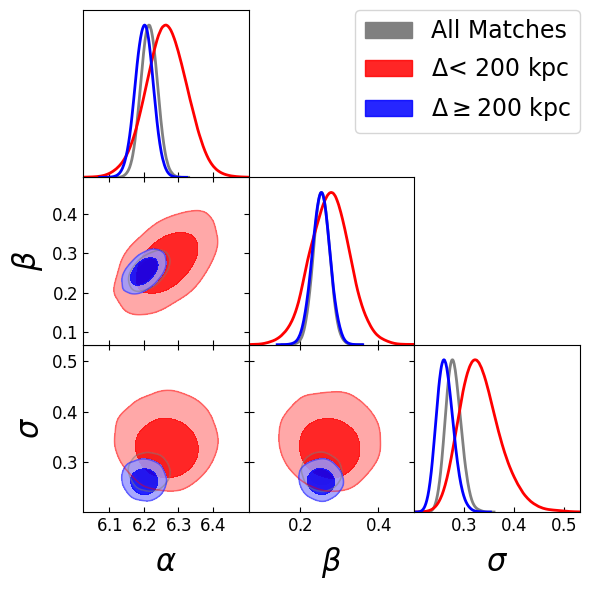}  
    \caption{}
    \label{fig:triangle_plot_195}
\end{subfigure}
\begin{subfigure}{.425\textwidth}
    \centering
    \includegraphics[width=.95\linewidth]{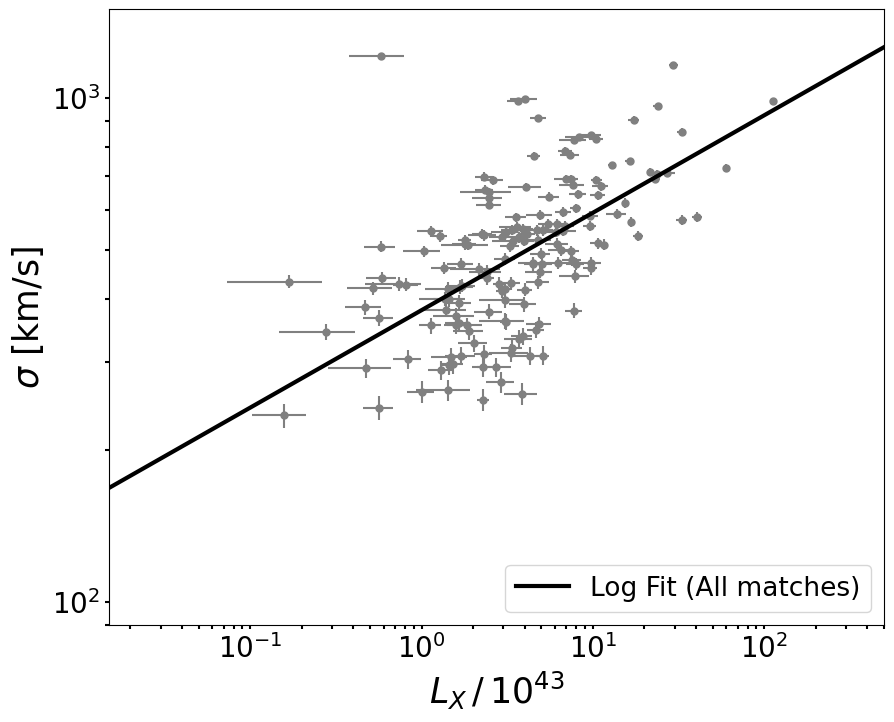}  
    \caption{}
    \label{fig:velD_Lx_cleaned}
\end{subfigure}
\begin{subfigure}{.35\textwidth}
    \centering
    \includegraphics[width=.95\linewidth]{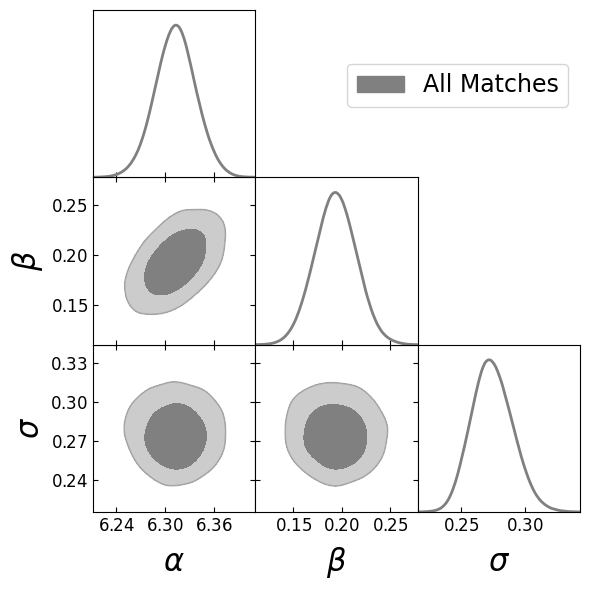}  
    \caption{}
    \label{fig:triangle_plot_195_cleaned}
\end{subfigure}
\caption{Left panels: Scaling relations between velocity dispersion ($\sigma$) and X-ray luminosity ($L_X$) for the optical/X-ray matches found using angular MHD cuts. {$\Delta$ is defined as the 2D distance between optical and X-ray centers for each match.} The power law fits for both subsets, as well as the full set of matches, were found via Monte Carlo simulation. For clarity, only the mean fits are plotted. The power law fits were calculated using natural log, but the figure axes are displayed in decimal log. For the top two panels on this side, the data is split between matches for which the distance between the optical and X-ray centers is less than ({red}) or greater than ({blue}) 200 kpc. (\ref{fig:velD_Lx_180}) Scaling relation for matches using the v180 volume-limited catalog. (\ref{fig:velD_Lx_195}) Scaling relation for matches using the v195 volume-limited catalog. (\ref{fig:velD_Lx_cleaned}) Scaling relation for matches using the cleaned version of the v195 volume-limited catalog, with the group membership being determined by the maximum MHD limit of the v195 catalog. Right panels: Comparison of key parameter posteriors from the Monte Carlo simulation used to find the power law fits for the scaling relations in the corresponding panels in the left column. Posteriors are show for the fit to all matches (gray), for matches whose X-ray and optical centers are less than 200 kpc away (red), and for matches with centers further apart than 200 kpc (blue)}
\label{fig:velD_Lx}
\end{figure*}

Using the matches determined by the MHD method, we can construct a scaling relation between the velocity dispersion of the optical galaxies and the X-ray luminosity. This scaling relation takes the form of the power law shown in equation \ref{eq:scaling_relation}.  Figures \ref{fig:velD_Lx_180} and \ref{fig:velD_Lx_195} present the scaling relation for the catalogs v180 and v195 from \cite{Tempel2014}, respectively. The posteriors of the power law fits are shown in figures \ref{fig:triangle_plot_180} and \ref{fig:triangle_plot_195}, respectively. For X-ray/optical matches with center distances greater than 200 kpc, the posteriors resemble those of the full catalog. However, matches with centers less than 200 kpc exhibit a slightly higher variance in the posteriors of the parameters. 

\begin{equation}
    \ln{\sigma\; \mathrm{km}^{-1} \mathrm{s} } = \beta \ln{\left(\frac{L_x}{10^{43}\; \mathrm{ergs\; s}^{-1}}\right) + N(\alpha,\sigma')}
    \label{eq:scaling_relation}
\end{equation}

Additionally, we applied the v195 MHD cutoff to each group in the complete {galaxy} catalog \cite{Tempel2014} as an alternative way to determine the membership of the cluster and recalculate the velocity dispersion measurements. Membership was determined by calculating the distance of each galaxy from the rest of the group and removing it if the distance exceeded the v195 MHD cutoff. We have recomputed the median value of right ascension and declination for the remaining group members and performed the cleaning for outliers using the {\it Clean} {algorithm} \citep{Mamon2013}.

\begin{table}

    \caption{Parameters of $\sigma-L_X$ scaling relation}
    \label{tab:lx_sig_params}
    \centering
    \begin{tabular*}{0.45\textwidth}
    {l@{\extracolsep{\fill}}llr}
        \hline\hline
        \multicolumn{1}{l}{Parameter} & \multicolumn{1}{l}{Mean} & \multicolumn{1}{c}{Standard Deviation}\\
        \hline
        \texttt{$\alpha_{v180}$} & 5.87 & 0.0545 \\
        \texttt{$\beta_{v180}$} & 0.141 & 0.0393 \\
        \texttt{$\sigma'_{\,v180}$} & 0.378 & 0.0316 \\
        \hline
        \texttt{$\alpha_{v195}$} & 6.01 & 0.0225 \\
        \texttt{$\beta_{v195}$} & 0.255 & 0.0212 \\
        \texttt{$\sigma'_{\,v195}$} & 0.279 & 0.016 \\
        \hline
        \texttt{$\alpha_{v195,\, cleaned}$} & 6.15 & 0.0231 \\
        \texttt{$\beta_{v195,\,cleaned}$} & 0.193 & 0.022 \\
        \texttt{$\sigma'_{\,v195,\, cleaned}$} & 0.275 & 0.0164 \\

        \hline
    \end{tabular*}
\end{table}

Replicating figures~\ref{fig:velD_Lx_180} and \ref{fig:velD_Lx_195} with these new, cleaned groups results in a scaling relation with a shallower slope and minimal change to the scatter compared to the uncleaned v195 scaling relation (Figure~\ref{fig:velD_Lx_cleaned}). Removing the outlier in the upper left corner of Fig.~\ref{fig:velD_Lx_cleaned} does not have any significant effect on the best power law fit parameters.

Notably, we see a significant increase in scatter between the scaling relations of v180 and v195, with v180 exhibiting a $2.4\sigma$ increase in scatter. 
The change in the level of scatter can be attributed to mass-dependent trends. While feedback does affect Lx, it does not affect the scatter on the velocity dispersion. Table \ref{tab:lx_sig_params} lists the scaling relation parameters for the set of all matches in the v180, v195, and cleaned v195 catalogs. Our lowest level of scatter is comparable to the expectations of increasing scatter toward low-z reported for galaxy clusters in \cite{Damsted23}.

\section{Conclusions}

Using a volume-limited optical group catalog obtained using a constant linking length and the contours of X-ray emission drawn at approximately the same level of baryonic overdensity, we built the procedure of source identification based on the MHD. Our identification method has superseded all previous methods of identifying the X-ray emission of galaxy groups, described by a 40-50\% completeness of identification of X-ray emission for groups with a velocity dispersion of 200$\,\mathrm{km}\,\mathrm{s}^{-1}$ and 70-80\% completeness for groups that have a dispersion of 300$\,\mathrm{km}\,\mathrm{s}^{-1}$.
We demonstrate that the newly added groups that have large separations between their X-ray and optical centers are characterized by the same scaling relations between the group's X-ray luminosity and the velocity dispersion of the group members, which rules out a difference in the feedback between those groups. Instead, the large distances are explained by over-merging in the optical group catalogs, and source confusion in X-rays, while only part of the X-ray emission or part of the optical structure match each other.

\begin{acknowledgements}
ET was funded by the Estonian Ministry of Education and Research (grant TK202), Estonian Research Council grant (PRG1006), and the European Union's Horizon Europe research and innovation programme (EXCOSM, grant No. 101159513).
\end{acknowledgements}

   \bibliographystyle{aa} 
   \bibliography{ref} 

\appendix
\onecolumn
\section{Catalogs of X-ray properties of the groups}

\begin{table*}

    \caption{Description of the columns of the v180 and v195 AXES-SDSS MHD catalogs.}
    \label{tab:catalogue_columns}
    \centering
    \begin{tabular*}{\textwidth}
    {l@{\extracolsep{\fill}}llr}
        \hline\hline
        \multicolumn{1}{l}{Column} & \multicolumn{1}{l}{Unit} & \multicolumn{1}{c}{Description} & \multicolumn{1}{r}{Example} \\
        \hline
        \texttt{GROUP\_ID} (1) & &  SDSS group  identification number from \cite{Tempel2014}& 424 \\
        \texttt{XC\_ID} (2) & & Extended X-ray contour ID & 931642001 \\
        \texttt{RA\_X} (3) & deg & Median right ascension of the matched part of X-ray contour (J2000) & 230.58942 \\
        \texttt{DEC\_X} (4) & deg & Median declination of the matched part of X-ray contour (J2000) & 
        +08.37373 \\
        \texttt{RA\_OPT} (5) & deg & Median right ascension of the matched part of the optical group (J2000) & 230.76036 \\
        \texttt{DEC\_OPT} (6) & deg & Median declination of the matched part of the optical group (J2000) & 
        +08.62381 \\
        \texttt{NMEM} (7) & & Number of spectroscopic members in SDSS group catalog  & 36 \\
        \texttt{ZSPEC} (8) & & SDSS group redshift & 0.036 \\
        \texttt{CLUVDISP\_GAP} (9) & km s$^{-1}$ & Gapper estimate of the cluster velocity dispersion & 705.4 \\
        \texttt{LX0124} (10) & ergs s$^{-1}$ & Luminosity in the (0.1-2.4) keV band of the cluster, aperture $R_\text{500c}$ & $1.2857\times10^{44}$ \\
        \texttt{ELX0124} (11) & ergs s$^{-1}$ & Uncertainty on \texttt{LX0124} & $5.9122\times10^{42}$ \\
        \texttt{FLUX052} (12) & ergs s$^{-1}$ cm$^{-2}$ & Galaxy cluster X-ray flux in the 0.5-2.0 keV band & $2.5118\times10^{-11}$ \\
        \texttt{EFLUX052} (13) & ergs s$^{-1}$ cm$^{-2}$ & Uncertainty on \texttt{FLUX052} & $1.1551\times10^{-12}$ \\
        \texttt{R\_E} (14) & arcmin & Apparent radial extent of X-ray emission at the contour level & 33.4127 \\
        \hline
    \end{tabular*}
\end{table*}

In Table \ref{tab:catalogue_columns} we describe the X-ray properties of the AXES-SDSS MHD Group catalogs. We release the X-ray properties of the v180 (Fig. \ref{fig:velD_Lx_180}) and v195 (Fig. \ref{fig:velD_Lx_195}) catalogs, identified using an MHD in megaparsecs. Source flux and luminosities are based on RASS data. In the flux extraction, we define polygon regions that follow the contours and define the effective radius (R\_E) as a circle with the area of the contour. As a source position in the released catalog, we provide a source ID and median coordinates of the contour. The properties of the source contain source ID, spectroscopic member counts, X-ray luminosity, redshift, and velocity dispersion of the group. The redshifts are reported in the CMB frame, using the catalogs of \cite{Tempel2014}. The catalogs described in Table \ref{tab:catalogue_columns} are only available in electronic form at the CDS.

\end{document}